\input epsf.tex
\documentclass [11pt]{article}
\Huge
\begin{document}
\renewcommand{\textheight}{590pt}
\begin{center}
	
\large
{\bf The description of an arbitrary configuration evolution
in the Conway's Game Life in terms of the elementary
configurations evolution in the first generation}

\vspace{0.2in}

Serguei~Vorojtsov\footnote{The current email: sv@phy.duke.edu}

\vspace{0.1in}

\normalsize
{\em Department of Theoretical Condensed Matter Physics\\
Institute for High Pressures Physics\\
Troitsk, Moscow region, Russia} \\

\end {center}
\vspace{0.3in}

\section*{Abstract}

The connection between the evolution of an arbitrary configuration and
the evolution of its parts in the first generation is
established. The equivalence of Conway's evolution rules to the
elementary configurations' (containing one, two, three, and four
pieces) evolution laws in the first generation has been proved.
\vspace{0.2in}

\section{Introduction}

Let us consider, for example, a two-dimensional configuration of pieces
shown in Fig.~1a. Figure~1b sets up a correspondence between the pieces
and the letters by which we denote each piece. Two cells in the plane are
neighboring ones if they have a common side or vertex. A piece in the cell
$C$, for instance, has three \mbox{adjacent pieces.}

\begin{figure} 
\epsfxsize=0.8\hsize 
\centerline{\epsffile{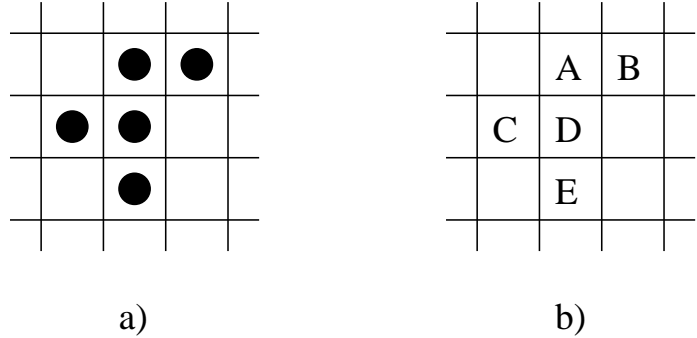}} 
\caption{Initial configuration.} 
\label{Fig1} 
\end{figure}

Conway's rules for a configuration evolution in the first generation are~[1]:

\begin{enumerate}

\item [1.]   {\bf Survival}: if a piece has two or three adjacent
pieces, it transfers to the next generation.

\item [2.]   {\bf Death}: otherwise a piece dies.

\item [3.]   {\bf Birth}: if an empty cell has exactly three
neighboring pieces, a piece is created in this cell in the next
generation.

\end{enumerate}

Rules 1, 2, and 3 are applied simultaneously and describe one "step". For
instance, Figure~2 illustrates one step (first generation) in the evolution of
the Fig.~1 configuration.

\begin{figure}
\epsfxsize=0.8\hsize
\centerline{\epsffile{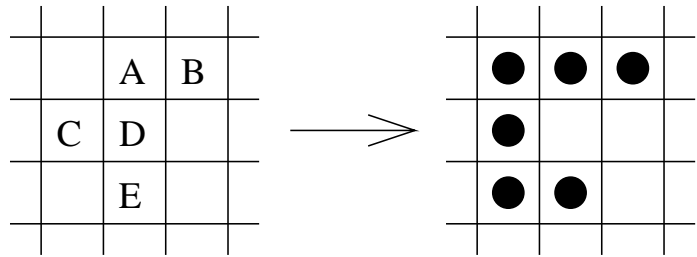}}
\caption{First evolution step for the initial configuration.}
\label{Fig2}
\end{figure}

\section{Definition of The Basic Concepts}

Let us discuss, for example, evolution of the configurations $a$ to $e$ shown
in Fig.~3, which are defined as {\em $1-$fragments} and can be obtained by
removing one of the pieces from the original configuration (Fig.~1). When
the piece $A$ is removed from the original configuration, we get the
initial fragment given by Fig.~3a, and so on.

\begin{figure}
\epsfxsize=0.8\hsize
\centerline{\epsffile{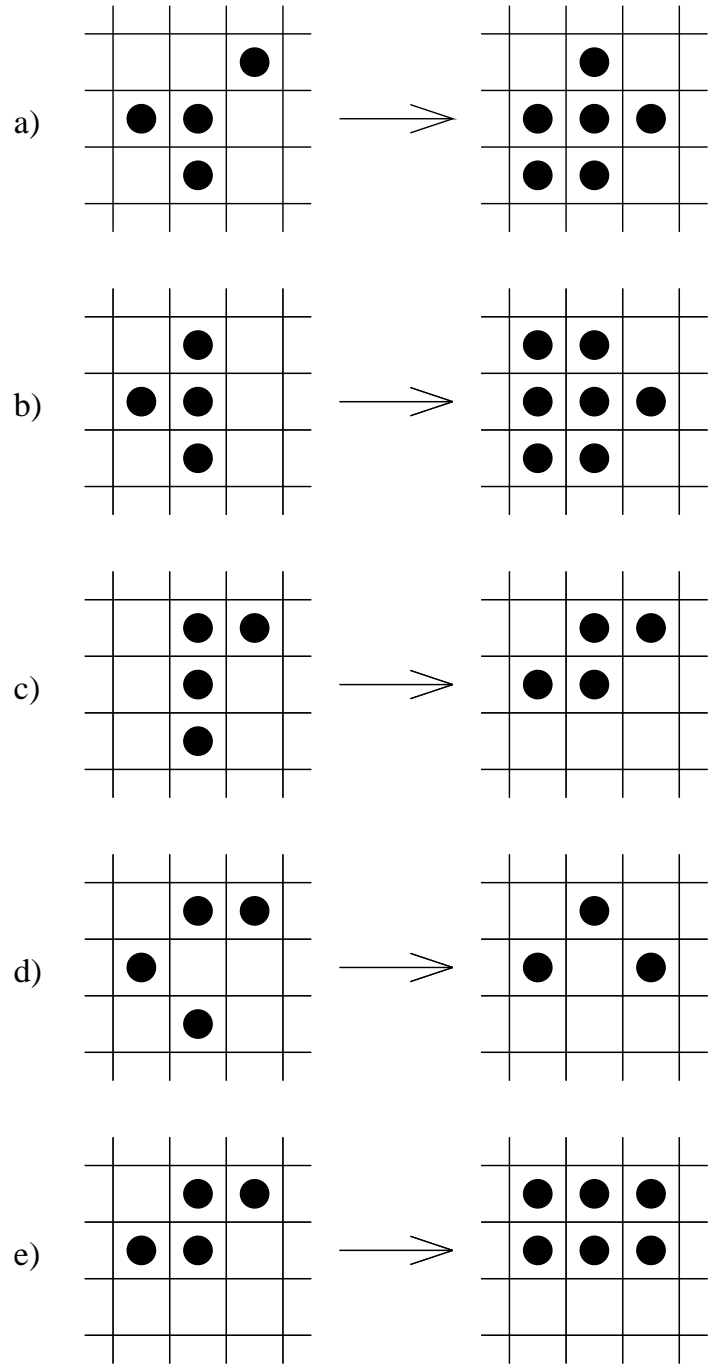}}
\caption{First evolution step of all 1-fragments of the initial configuration.}
\label{Fig3}
\end{figure}

Now let all the patterns evolve in the first generation and count the
number of pieces in each cell. These numbers are given in Fig.~4.

\begin{figure}
\epsfxsize=0.35\hsize
\centerline{\epsffile{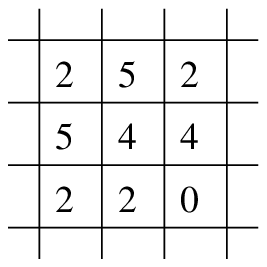}}
\caption{Values of S(F) for 1-fragments in the first generation.}
\label{Fig4}
\end{figure}

The number of pieces in cell $F$ is denoted by $S(F)$. One can see that 
the first generation of the Fig.~1 configuration shown in Fig.~2 can be
obtained by putting pieces to the cells with $S=2$ and 5, and leaving empty
the cells corresponding to $S=0$ and 4. Thereby we have reduced the evolution
of the configuration containing five pieces to the evolution of the patterns
containing four pieces (Fig.~3). We call this procedure {\em $5\to
4$ reduction}.

\section {$N\to N-1$~Reduction Theorem (Removal of One Piece)}

Suppose that the initial pattern is composed of $N$ pieces. Let us
consider the first generation evolution of its $1-$fragments which contain
$N-1$ pieces each and establish the correspondence between the presence or
absence of a piece in the cell in the first generation and the value of $S(F)$
\mbox{in this cell.}

\begin{enumerate}

\item [1.]   {\bf Survival}:

\begin{enumerate}

\item [(a)] if a piece $F$ has exactly two pieces in its neighboring cells then
this cell is filled in the first generations of all $1-$fragments except for
the three fragments obtained by removing from the initial pattern either the
piece $F$ or one of its neighbors, i.e., $S(F)=N-3$;

\item [(b)] if a piece $F$ has three pieces in its adjacent cells then its
cell is filled in the first generations of all $1-$fragments, i.e.,
$S(F)=N$.

\end{enumerate}

\item [2.]   {\bf Death}:

\begin{enumerate}

\item [(a)] for any number of pieces in the cells adjacent to a piece $F$
except 4 (i.e., at 0, 1, 5, 6, 7, and 8), its cell is empty in the first
generations of all $1-$fragments, hence $S(F)=0$;

\item [(b)] if a piece $F$ has four neighbors then it is present in the first
generation of four $1-$fragments obtained by removing from the initial
pattern one of the four pieces adjacent to $F$, thus in this case
$S(F)=4$.

\end{enumerate}

\item [3.]   {\bf Creation}:

\begin{enumerate}

\item [(a)] if an empty cell $F$ in the initial configuration has three
neighboring pieces then a new piece is generated in this cell in the
first evolution step of all $1-$fragments, except for the three patterns
obtained by removing one of these three pieces, i.e., $S(F)=N-3$;

\item [(b)] if an empty cell $F$ has four pieces in its adjacent cells then
$F=4$;

\item [(c)] at any other number of pieces in the cells adjacent to an empty
cell, a piece cannot be generated in this cell in the first generation
of all $1-$fragments, hence $S(F)=0$.

\end{enumerate}

\end{enumerate}

These results are summarized in the convenient form in Table~1. This table
clearly shows that the numbers $S(F)=N-3$ and $N$ correspond to the presence
of a piece in the cell, whereas $S(F)=0$ and 4 corresponds to the absence of a
piece.

\bigskip

\begin{tabular}{l|c|c|c}
Evolution & The $\#$ of neighboring & $S(F)$ & $S(F)$ \\
process& cell $F$ pieces at which & corresponding to & corresponding to \\
& the process takes place & $1-$~fragments & $2-$~fragments \\
\hline
1. Survival & 2 & $N-3$ & $(N-3)(N-4)/2$ \\
             & 3 & $N$ & $(N+3)(N-4)/2$ \\
\hline
2. Death & $0,1,6,7,8$ & $0$ & $0$ \\
              & $4$ & $4$ & $4(N-4)+6$ \\
              & $5$ & $0$ & $10$ \\
\hline
3. Birth & $3$ & $N-3$ & $(N-3)(N-4)/2$ \\
\hline
4. Absence     & $0,1,2,6,7,8$ & $0$ & $0$ \\
   of birth    & $4$ & $4$ & $4(N-4)$ \\
               & $5$ & $0$ & $10$ \\
\hline
\end{tabular}

\bigskip

{Table 1. The rules of correspondence between $S(F)$ and the presence or
absence of a piece in the cell $F$ for $N \to N-1$ and $N \to N-2$~reductions.}

\bigskip

At the same time, if $N-3=0$, or $N-3=4$, or $N=4$, i.e. $N=3$, 4, or 7,
we cannot establish one-to-one correspondence between the values of $S(F)$
and the presence or absence of a piece in the cell $F$ in the first
generation. Configurations containing these numbers of pieces we name
{\em $1-$irreducible} (i.e., they cannot be analyzed by removing one piece).

Thus, the evolution of an arbitrarily complex configuration of more than seven
pieces can be analyzed on the base of the first generation evolution rules
for various configurations with seven pieces and the correspondence rules
summarized in the Table~1.

Now we can formulate the $N\to N-1$ reduction theorem. {\em Conway's
evolution rules are equivalent to the combination of the rules for one
evolution step of $1-$irreducible configurations (i.e., all configurations
composed of $N=3$, $4$, and $7$ pieces - whereas the fact that configurations
with $N=1$ and $2$ annihilate after the first step is taken for granted) and
the correspondence rules formulated as follows: if $S(F)=N-3$ or $N$, a
piece is generated (conserved) in the cell $F$, otherwise cell $F$ must be
empty after the first step}.
\vskip 15pt

\section{$N \to N-M$ Reduction Theorem (Elimination of M Pieces)}

Alongside with the original configuration composed of $N$ pieces, let us
consider one evolution step of $C_N^M$ (number of the combinations where one
can  choose $M$ of $N$) $M-$fragments containing $N-M$ pieces each.

The rules of correspondence for the general case of $M-$~fragments are
summarized in Table~2. These rules can be epitomized in the following
statement: if $S(F)=C^M_{N-3}$ or $C_{N-4}^{M-1}$ then there is a piece in
the cell $F$ after one evolution step, otherwise the cell is vacant (it is
obvious that the rule applies only to $M-$ reducible configurations). It is
assumed in Table~2 that $C_I^K=0$ if $I<0$ and/or $K<0$.

\bigskip

\begin{tabular}{l|c|c|c}
Evolution & The $\#$ of neighb. & $S(F)$ & $S(F)$ \\
process & $F$ pieces at which & corresponding to & corresponding to
\\ & the process occurs & $M-$"fragments" & $M=N-4$ \\
\hline
1. Survival & $2$ & $C_3^0C_{N-3}^M$ & $N-3$ \\
& $3$ & $C_4^0C_{N-4}^M+C_4^1C_{N-4}^{M-1}$ & $1+4(N-4)$ \\
\hline
2. Death & $0,1$ & $0$ & $0$ \\
& $L=4,5,\ldots ,8$ & $C_L^{L-3}C_{N-L}^{M-(L-3)}+$ & $L(L-1)/6$ \\
& & $C_L^{L-2}C_{N-L-1}^{M-(L-2)}$ & $\left[ (L+1)(N-L)-3\right]$ \\
\hline
3. Birth & $3$ & $C_3^0C_{N-3}^M$ & $N-3$ \\
\hline
4. Absence     & $0,1,2$ & $0$ & $0$ \\
of birth & $L=4,5,\ldots ,8$ & $C_L^{L-3}C_{N-L}^{M-(L-3)}$
& $L(L-1)(L-2)$ \\
& & & $(N-L)/6$ \\
\hline
\end{tabular}

\bigskip

{Table 2. The same as in Table 1 for the general case of $N\to N-M$ 
reduction and $N\to 4$~reduction}.

\bigskip

Examples of the reasonings according to which Table~2 has been composed
are given below.

\begin{enumerate}

\item [A)]   {\bf Survival}. Suppose that a piece in the cell $F$ has three
neighbors. Let us select a subconfiguration of the initial configuration
that includes a piece in the cell $F$ and its neighboring pieces. This piece
survives the next evolution step in two cases only. Firstly, when we
remove only pieces not included in the subconfiguration then the
resulting number of $M-$fragments is $C_4^0C_{N-4}^M$. Secondly, when we
remove only one piece of the subconfiguration then the number of these $M-$
fragments is $C_4^1C_{N-4}^{M-1}$.

\item [B)]   {\bf Death}. Suppose that a piece in the cell $F$ has four
neighbors. In this case, we include into subconfiguration only the pieces
adjacent to $F$. A piece $F$ will remain only in the following cases:

\begin{enumerate}

\item [(1)]   One of the pieces in the subconfiguration is removed. The
number of such $M-$fragments is $C_4^1C_{N-4}^{M-1}$.

\item [(2)]   Two pieces included in the configuration are removed and
a piece $F$ is saved. In this case the number of $M-$fragments is
$C_4^2C_{N-5}^{M-2}$.

\end{enumerate}

\end{enumerate}

The correspondence rules for $2-$fragments are given in Table~1. By
equating, as previously, the values of $S(F)$ corresponding to the presence
and the absence of a piece at the cell $F$, we come to the conclusion that the
configurations of $N=3$, 4, 5, 8, and 11 pieces are $2-$irreducible (they
cannot be analyzed by removing two pieces). An important point, however, is
that the configurations containing seven pieces are $2-$reducible.

Thus, despite the fact that the configurations with $N=7$ are
$1-$irreducible they turn out to be $2-$reducible (a piece occupies the cell
$F$ if, and only if, $S(F)=6$ or 15). It means that we do not have to know the
evolution laws for the configurations of seven pieces which were necessary
according to the $N\to N-1$ reduction theorem.

Since all configurations with $N>4$ are $(N-4)-$reducible (as follows from
Table~2), $N\to 4$ reduction theorem has, probably, the most
rational form. The correspondence rules for this case are listed in Table~2.

\bigskip\noindent

{\bf $N-4$ reduction theorem}: {\em Conway's three
evolution rules are equivalent to the postulated evolution laws of all
configurations containing $N=3$ and $4$ pieces in the first generation 
(configurations with $N=1,2$ annihilate after the first step)
and the rule of correspondence: a piece is generated or conserved at the
cell $F$ if, and only if, $S(F)=N-3$ or $1+4(N-4)$.} 

\bigskip

This theorem prescribes how to reduce the evolution of an arbitrarily complex
configuration to the evolution of the elementary configurations containing
$N=4$ in the first generation.

\section{Discussion of The Results}

Conway's rules are expressed in the local form. Indeed, the presence or
absence of a piece, say, in the cell $D$ in the first generation of the
configuration $A...E$ (Fig.~2) is determined exclusively by the state of the
adjacent cells. On the contrary, the rules based in the discussed theorem have
the global form because they are based on the laws of evolution of the
configurations containing four pieces separated by an arbitrary distance.

Physically, one can say that the fundamental "interaction" in the Conway's game
Life is the four-piece "interaction". Here we consider three-piece
"interaction" as a particular case of the four-piece "interaction" when one of
the pieces is well separated from the other three (we also assume that one- and
two-piece "interactions" vanish in the first generation). 

Finally, we note that our theorems have been formulated strictly in accordance
with the Conway's evolution rules. It is possible to formulate similar global
rules for any cellular automata.

\section*{Acknowledgements}

The author is grateful to M.~Savrov for the discussion of the results,
I.~Beloborodov for the help in the work, Prof. S.~Gordyunin for
the support, and Brendan Crowley for his proof-reading and zest for life.

\section*{References} 
[1] M.~Gardner, {\em The Unexpected Hanging and Other Mathematical
Diversions}, Simon and Schuster, New York (1969). \\

\end{document}